# VARIOUS MODELS OF PROCESS OF THE LEARNING, BASED ON THE NUMERICAL SOLUTION OF THE DIFFERENTIAL EQUATIONS

Mayer Robert Valerievich, http://maier-rv.glazov.net
The Glazov Korolenko State Pedagogical Institute, Glazov, Russia,

The principles on which can be based computer model of process of training are formulated. Are considered: 1) the unicomponent model, which is recognizing that educational information consists of equal elements; 2) the multicomponent model, which is considering that knowledge is assimilate with a various strength, and on lesson weak knowledge becomes strong; 3) the generalized multicomponent model which considers change of working capacity of the pupil and various complexity of studied elements of a training material. Typical results of imitating modeling of learning process are presented in article.

# РАЗЛИЧНЫЕ МОДЕЛИ ПРОЦЕССА ОБУЧЕНИЯ, ОСНОВАННЫЕ НА ЧИСЛЕННОМ РЕШЕНИИ ДИФФЕРЕНЦИАЛЬНЫХ УРАВНЕНИЙ

Майер Роберт Валерьевич, http://maier-rv.glazov.net
Глазовский государственный педагогический институт, Глазов, Россия

Одна из проблем дидактики состоит в следующем: как, зная параметры ученика, его начальный уровень знаний и воздействие, оказываемое учителем, предсказать количество знаний ученика в последующие моменты времени [1, 3]. Метод имитационного моделирования [4] позволяет создать компьютерную программу, симулирующую поведение системы "учитель–ученик", и исследовать влияние ее параметров на результаты обучения.

**1. Основные принципы моделирования.** Сформулируем принципы, которые могут быть положены в основу компьютерной модели обучения:

1. Скорость изменения количества знаний равна сумме скорости усвоения и скорости забывания.

2. Обучение организовано так, что ученик хотя бы в течение нескольких минут удерживает в памяти каждый элемент учебного материала (ЭУМ) и может его повторить. При этом учащийся стремится запомнить (пусть не на долго) всю сообщаемую ему информацию $Z_0$. Уровень требований учителя U равен количеству сообщаемых учителем знаний $Z_0$.

3. Скорость увеличения знаний пропорциональна: 1) количестве знаний Z ученика в степени b (b из интервала [0; 1]); 2) мотивации M или количеству усилий F, затрачиваемых учеником. Действительно, чем больше ученик знает, тем легче он усваивает новые знания из–за образующихся ассоциативных связей с имеющимися.

С другой стороны, чем ниже мотивация М учащегося, тем меньше усилий $F$ он затрачивает и тем ниже скорость увеличения знаний. Если прирост знаний много меньше их общего количества Z (обучение в течение одного или нескольких занятий), то можно считать, что Z остается постоянным и b = 0.

5. Усилия ученика F (мотивация М к учебной деятельности) прямо пропорциональна разности между уровнем предъявляемых требований U и уровнем знаний Z: F = M = k(U – Z). В случае, когда U – Z превышает некоторое пороговое значение C, ученик перестает прикладывать усилия: F = M = 0.

6. Скорость забывания пропорциональна количеству имеющихся у учащегося знаний: dZ/dt = – $\gamma$ Z, ($\gamma$ > 0), где $\gamma$ – коэффициент забывания.

**2. Однокомпонентная модель обучения.** В простейшем случае можно считать, что сообщаемая учителем информация (знания) является совокупностью равноправных несвязанных между собой элементов, число которых пропорционально ее количеству Z. Все элементы учебного материала (ЭУМ) одинаково легко запоминаются и с одинаковой скоростью забываются. В этом случае процесс обучения можно описать уравнением:

$$\frac{dZ}{dt} = \begin{cases} \alpha Z^b (U - Z) - \gamma Z, & U \leq Z + C, \\ -\gamma Z, & U > Z + C. \end{cases}$$

Здесь $Z$ – количество знаний ученика, $U$ – уровень требований учителя, равный количеству сообщенных им знаний $Z_0$, $\alpha$ – коэффициент усвоения ученика, $\gamma$ – коэффициент забывания, $C$ – критическое значение.

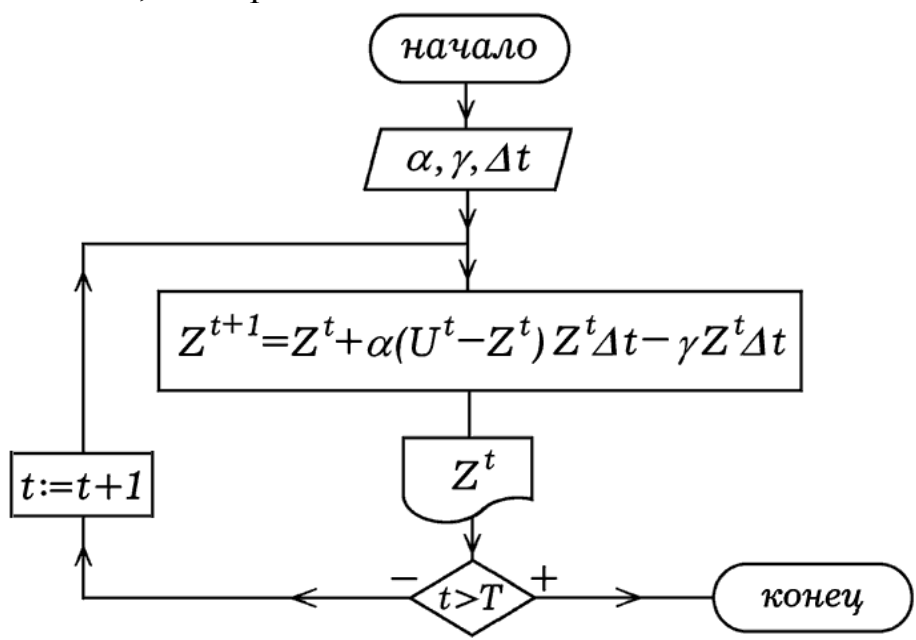

Рис. 1. Алгоритм имитационного моделирования процесса обучения.

Когда Z мало, скорость роста уровня знаний невысока из–за отсутствия возможности образования ассоциативных связей. По мере увеличения Z она растет, но при Z стремящемся к U уменьшается за счет снижения усилий F (мотивации M). Если U превышает Z на величину большую критического значения C, то ученик перестает учиться.

Чтобы промоделировать процесс обучения, необходимо перейти от записанного выше диффуравнения к конечно–разностному уравнению [2, с. 55–56]. Используемая компьютерная программа содержит цикл по времени t, в котором вычисляется скорость увеличения знаний, определяется уровень знаний в следующий дискретный момент времени t+1, строится соответствующая точка графика, после чего все повторяется снова. Упрощенная блок–схема алгоритма представлена на рис. 1.

Используя компьютерную модель обучения, можно обосновать известный принцип "от простого к сложному". Допустим, сначала изучается сложная тема, а затем простая, то есть сначала уровень требований учителя высокий, а затем низкий ($U_1 > U_2$). Если $U_1$ очень сильно превосходит уровень знаний Z ученика, то мотивация к обучению пропадает, и уровень знаний не растет (ученик просто не может усвоить материал). Если же $U_1 < Z + C$, то ученик усваивает сложную тему, прилагая большое количество усилий F. При изучении второй более простой темы скорость роста знаний невысока из–за того, что уровень требований $U_2$ незначительно превосходит уровень знаний Z ученика, и он не затрачивает много усилий F. В идеале при изучении различных тем ученик должен затрачивать примерно одинаковое количество усилий.

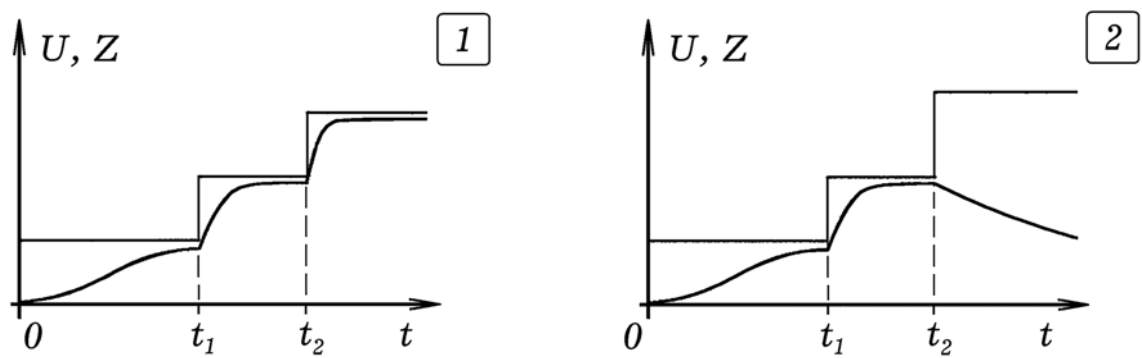

Рис. 2. Обучение при скачкообразном повышении уровня требований.

На рис. 2 показано, как ведет себя рассмотренная выше модель обучения, когда уровень U предъявляемых требований (количество изучаемого материала, сложность заданий) скачкообразно увеличивается. Сначала учащимся предлагают сравнительно простые задания; когда они их освоят, дают задания сложнее, затем еще сложнее и т.д. Для того, чтобы уровень знаний рос, необходимо обеспечить не очень большой разрыв между Z и U (рис. 2.1). Слишком резкое увеличение уровня требований (сложности изучаемого материала) приводит к снижению мотивации и уменьшению уровня знания вследствие забывания (рис. 2.2). Если сначала предло-жить сложные задания (уровень требований U высок), а затем простые, то обучения происходить не будет. Для повышения эффективности обучения необходимо таким образом подбирать уровень требований (сложность предлагаемых учащимся заданий), чтобы: 1) сохранялась высокая мотивация к обучению; 2) ученик при изучении различных тем работал бы с одинаковым напряжением, прилагая примерно равное количество усилий; 3) работа, совершаемая в течение занятия, не превышала бы некоторое пороговое значение.

**3. Учет изменения работоспособности ученика.** Будем считать, что скорость увеличения знаний ученика пропорциональная его коэффициенту научения a, работоспособности r, приложенным усилиям F (мотивации M) и уровню знаний Z в степени b. Работоспособность r зависит от степени усталости ученика; она сначала равна $r_0$, а затем по мере совершения учеником работы P плавно снижается до 0. Получаем следующую математическую модель [2, с. 66 – 68]:

$$dZ/dt = r\alpha F Z^b - \gamma Z, \quad r = r_0/(1+\exp(k_1(P-P_0))), \quad (0 < r_0 \leq 1).$$

Здесь $P_0$ — работа, совершаемая учеником на занятии (решение задач, выполнение заданий), после выполнения которой его работоспособность уменьшается от $r_0 = 1$ до 0,5. При обучении уровень требований учителя (сообщаемые им знания) больше уровня знаний ученика (U > Z), и учебная работа, совершенная учеником (число выполненных заданий), зависит от приложенных усилий (интенсивности мыслительной деятельности) и длительности обучения. Усилия ученика F пропорциональны его мотивации или разности между уровнем требований U учителя и количеством знаний Z:

$$F = U - Z, \quad \Delta P = k_2 F \Delta t = k_2(U-Z)\Delta t, \quad P = k_2 \sum_{i=1}^{N} F_i \Delta t.$$

Здесь N — число элементарных промежутков времени, на которое разбит урок. Если уровень предъявляемых требований мал (U < Z), то есть ученик на уроке занят решением простых для него задач, то затрачиваемые им усилия пропорциональны времени: P = k t. Это позволяет учесть появление у ученика усталости и снижение работоспособности даже в случае, когда он выполняет простые задания длительное время. В перерывах между занятиями ученики отдыхают, работоспособность восстанавливается. Максимальная работоспособность ученика в данное время учебного дня снижается по экспоненциальному закону. Получаем уравнения:

$$dr/dt = k_3(r_{\max} - r),$$
$$r(t) = r_{\max} - (r_{\max} - r_0)\exp(-k_3(t-t_0)),$$
$$r_{\max} = \exp(-k_4 t).$$

Здесь $r_0 = r(t_0)$ — работоспособность в момент начала отдыха $t_0$ (то есть в конце урока), где $r_{\max}$ — максимальная работоспособность ученика в данное время $t$ учебного дня. Скорость увеличения знаний при прочих равных условиях тем выше, чем меньше субъективная сложность (трудность понимания) S изучаемого материала. Сложность учебного материала S лежит в интервале от 0 до 1 и в общем случае зависит от уровня изучения других вопросов. Математическая модель выражается уравнениями:

Во время обучения ($U > Z$): $\dfrac{dZ}{dt} = \dfrac{(1-S)\alpha F Z^b}{1+\exp(k(P-P_0))} - \gamma Z$,

Во время перерыва ($U = 0$): $dZ/dt = -\gamma Z$.

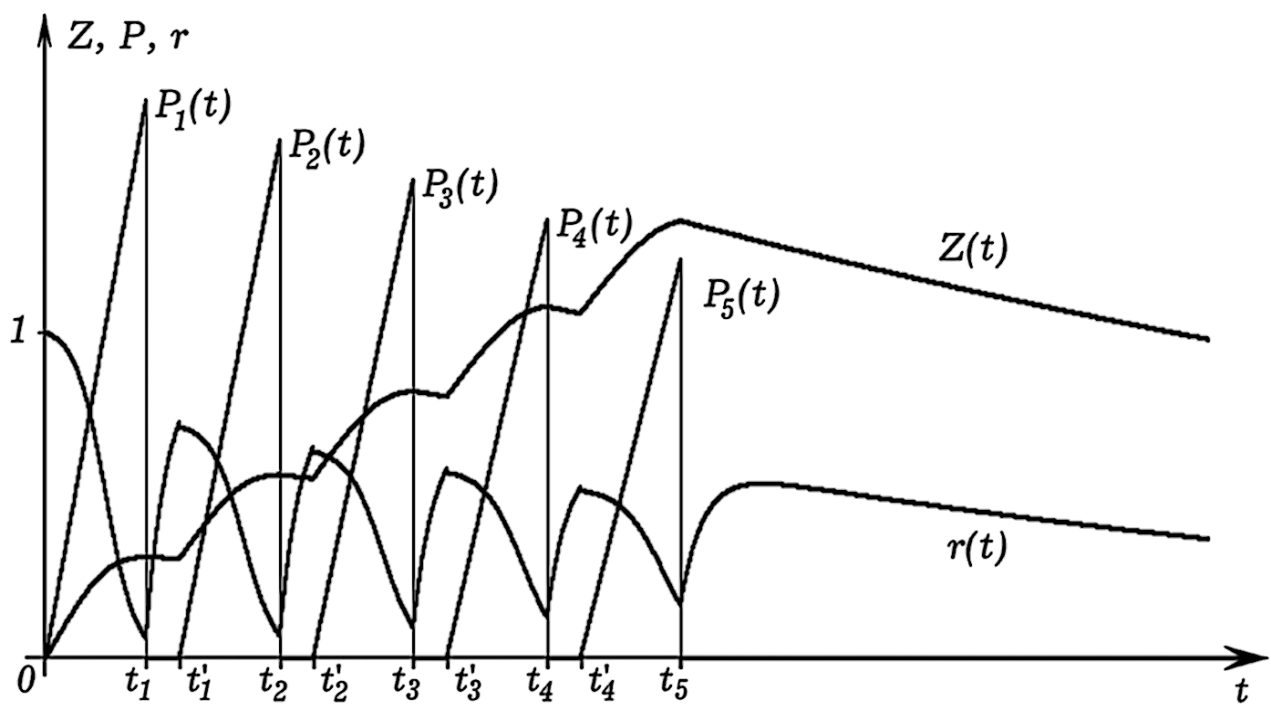

Рис. 3. Модель, учитывающая изменения работоспособности ученика.

### 4. Многокомпонентная модель процесса обучения.

Выше предполагалось, что все элементы учебного материала усваиваются одинаково прочно. Но на практике те знания, которые включены в учебную деятельность ученика, запоминаются значительно прочнее, чем знания, которые он не использует. При этом формируются интеллектуальные умения и навыки. Можно предположить, что компьютерная имитация будет более точно соответствовать реальному процессу обучения, если учесть, что: 1) прочность усвоения различных ЭУМ неодинакова, поэтому все ЭУМ следует разделить на несколько категорий; 2) прочные знания забываются существенно медленнее непрочных; 3) непрочные знания при их использовании учащимся постепенно становятся прочными [2, с. 70 – 72]. Предлагаемая многокомпонентная модель обучения выражается системой уравнений:

$$dZ_1/dt = r(1-S)(\alpha_1 F Z^b - \alpha_2 Z_1) - \gamma_1 Z_1,$$
$$dZ_2/dt = r(1-S)(\alpha_2 Z_1 - \alpha_3 Z_2) - \gamma_2 Z_2,$$
$$..., \ dZ_n/dt = r(1-S)\alpha_n Z_{n-1} - \gamma_n Z_n,$$
$$Z = Z_1 + Z_2 + Z_3 + ... + Z_n, \ F = U - Z. \quad 6$$

Здесь $U$ — уровень требований, предъявляемый учителем, равный сообщаемым знаниям $Z_0$, $Z$ — суммарные знания, $Z_1$ — количество самых непрочных знаний первой категории с высоким коэффициентом забывания $\gamma_1$, а $Z_n$ — количество самых прочных знания $n$–ой категории с низким $\gamma_n$.

Коэффициенты усвоения $\alpha_i$ характеризуют быстроту перехода знаний (i – 1)–ой категории в знания i–ой категории. Если прирост знаний ученика существенно меньше их общего количества, то b = 0. При изучении одной темы растет общее количество знаний Z, и постепенно увеличивается количество прочных знаний $Z_4$.

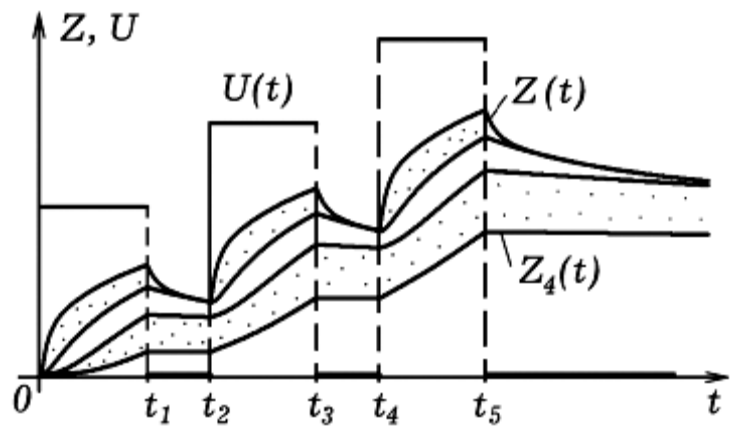

Рис. 4. Результаты использования четырех-компонентной модели обучения.

**4. Обобщенная многокомпонентная модель обучения**

Автором предложена обобщенная модель обучения, не имеющая аналогов в известной ему литературе. Пусть работоспособность ученика в начале учебного дня $r_0 = 1$. В любой момент времени $Z(t) = Z_1(t) + ... + Z_n(t)$.

Во время обучения:

$$F = U - Z > 0, \ r = r_0/(1 + \exp(k_1(P - P_0))), \ P = k_2 \int_{t_0}^{t}(1+S)(U-Z)dt,$$

$$dZ_1/dt = r(1-S)(\alpha_1 F Z^b - \alpha_2 Z_1) - \gamma_1 Z_1,$$
$$dZ_2/dt = r(1-S)(\alpha_2 Z_1 - \alpha_3 Z_2) - \gamma_2 Z_2,$$
$$..., \ dZ_n/dt = r(1-S)\alpha_n Z_{n-1} - \gamma_n Z_n.$$

Время перерыва: $U = 0$, $dr/dt = k_3(r_{\max} - r)$, $r_{\max} = \exp(-k_4 t)$,

$$dZ_1/dt = -\gamma_1 Z_1, \ dZ_2/dt = -\gamma_2 Z_2, \ ..., \ dZ_n/dt = -\gamma_n Z_n.$$

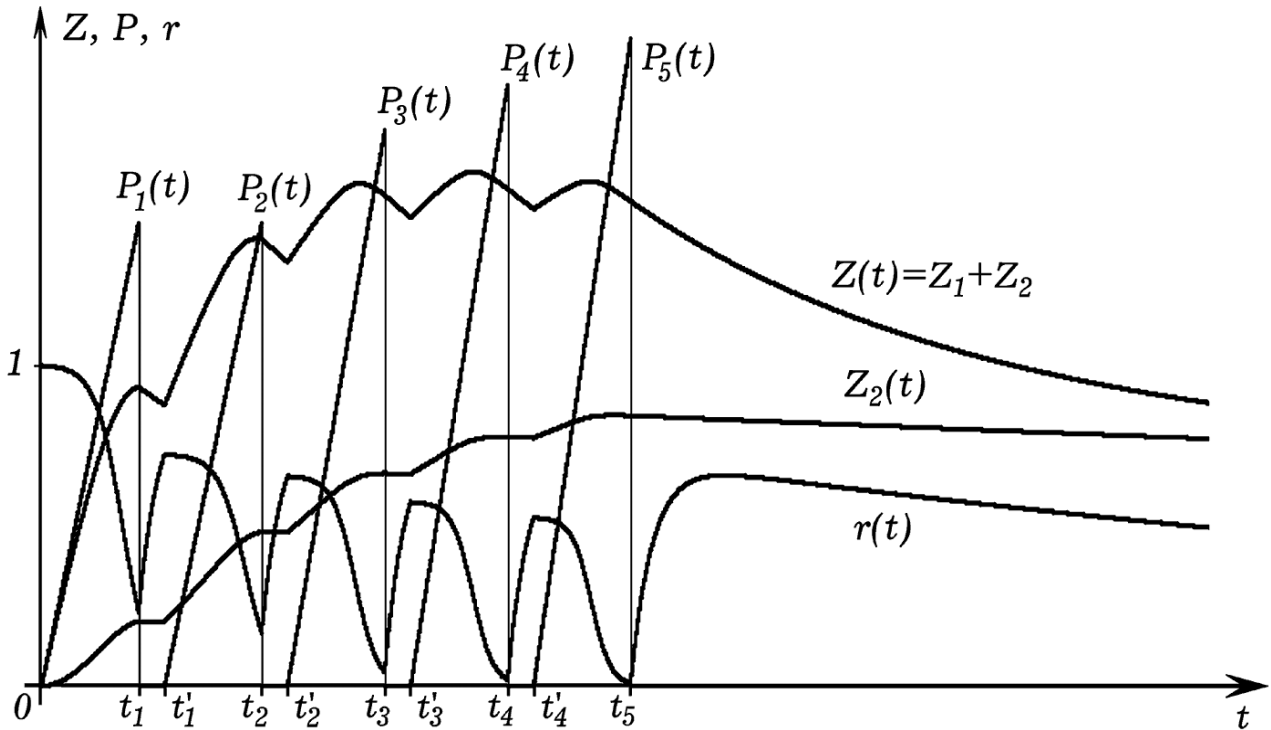

Рис. 5. Двухкомпонентная модель, учитывающая изменения работоспособности ученика.

Результаты использования двухкомпонентной модели приведены на рис. 5. Прочные знания $Z_2$ в процессе обучения растут, а после — практически не забываются. Непрочные знания $Z_1 = Z - Z_2$ забываются существенно быстрее. Работоспособность ученика во время урока плавно снижается, а во время перерывов — повышается до величины, которая постепенно уменьшается в течение дня из–за накапливающейся усталости.

Рассмотренные выше имитационные модели учебного процесса позволяют создать обучающую программу (пакет программ), моделирующую обучение школьников, которую можно использовать для тренировки студентов педагогических вузов. Она должна допускать изменение параметров учеников, длительность занятий, распределения учебного материала и стратегии поведения учителя. В процессе ее работы студент ("учитель") изменяет скорость подачи учебной информации, быстро реагирует на вопросы учеников, проводит контрольные работы, ставит оценки, пытаясь добиться наибольшего уровня знаний за заданное время. После окончания обучения на экран выводятся графики, показывающие изменение знаний "учеников" класса, обучающая программа анализирует работу "учителя" и ставит ему оценку.

## Литература